\documentclass[a4paper,preprint,onecolumn,showpacs,amsmath]{revtex4}

\usepackage{graphicx}
\usepackage[dvips]{color}

\begin{document}
\title{Two Energy Scales and two Quasiparticle Dynamics
in the Superconducting State of Underdoped Cuprates}

\author{M. Le Tacon, A. Sacuto}
\affiliation{Laboratoire Mat\'eriaux et Ph\'enom$\grave{e}$nes Quantiques (UMR 7162 CNRS),
Universit\'e Paris 7, 2 place Jussieu
75251 Paris, France}
\affiliation{Laboratoire de Physique du Solide, ESPCI, 10 rue Vauquelin 75231 Paris, France}
\author{A. Georges}
\affiliation{Centre de Physique Th\'eorique, Ecole Polytechnique, 91128 Palaiseau Cedex, France}
\author{G. Kotliar}
\affiliation{Centre de Physique Th\'eorique, Ecole Polytechnique, 91128 Palaiseau Cedex, France}
\affiliation{Serin Physics Laboratory, Rutgers University, USA}
\author{Y. Gallais}
\affiliation{Departments of Physics and Applied Physics, Columbia University New York, NY 10027, USA}
\author{D. Colson, A. Forget}
\affiliation{Service de Physique de l'Etat Condens\'{e}, CEA-Saclay, 91191 Gif-sur-Yvette, France}

\date{\today}

\begin{abstract}
The superconducting state of underdoped cuprates
is often described in terms of a single energy-scale,
associated with the maximum of the (\textit{d}-wave) gap.
Here, we report on electronic
Raman scattering results, which show that the gap function
in the underdoped regime is 
characterized by two energy scales, depending on
doping in opposite manners. 
Their ratios to the maximum
critical temperature are found to be universal in cuprates.
Our experimental results also reveal
two different quasiparticle dynamics in the
underdoped superconducting state,
associated with two regions of momentum space: nodal
regions near the zeros
of the
gap and antinodal regions. While antinodal 
quasiparticles quickly loose coherence as doping is reduced, coherent nodal
quasiparticles persist down to low doping levels. A theoretical
analysis using a new sum-rule allows us to relate
the low-frequency-dependence of the Raman response to the
temperature-dependence of the superfluid density, both controlled
by nodal excitations.
\end{abstract}


\maketitle
\date{\today}
Understanding the origin of high temperature superconductivity in copper oxide
based materials is one of the outstanding problems in modern condensed
 matter theory. Two decades of intensive theoretical and experimental
 work have revealed that this phenomenon takes place in various families of
 cuprates, all containing copper oxide layers separated by building
 blocks that provide a reservoir of carriers. The cuprates phase diagram is
 remarkably universal. When there is nominally one hole per copper
 in the copper oxygen planes the cuprates are in an insulating phase due to
 correlations known as a Mott insulator. As the number of charge
carriers is increased, a process called doping, the cuprates turn into a
 \textit{d}-wave superconductor.

The superconducting temperature has a dome like shape as a function of
doping and cuprates exhibit two physically distinct regimes: i) the
under-doped regime where the critical temperature $T_c$ increases with
doping, ii) an over-doped regime where $T_c$ decreases with doping. They are
separated by optimal doping where $T_c$ reaches its maximum. There is now
consensus that the "normal state" above the superconducting critical
temperature in the under-doped and optimally doped regimes is far
from being normal. In particular, a pseudo gap develops in the
underdoped regime, corresponding to a partial suppression of spin and
charge excitations. This phenomenon lies outside the standard theory of solids and
manifests itself in all experiments, such as neutron scattering
cross sections, nuclear magnetic resonance (NMR), magnetic
susceptibility, specific heat, angle resolved photoemission spectroscopy
(ARPES), optical conductivity, tunneling and Raman spectroscopies (for a
review, see Ref.~\cite{Timusk_RPP99}).

%
In contrast, there is no consensus on whether the superconducting state in the underdoped
regime, which emerges from the Mott insulating state, can be described by
the standard BCS-Migdal-Eliashberg theory for a \textit{d}-wave superconductor.
In this article, we address this issue using Electronic Raman Spectroscopy(ERS).
 ERS is a powerful tool for probing quasiparticles of the superconducting state
 in selected parts of the momentum space namely the antinodal (ANR) and nodal
 (NR) regions where the amplitude of the superconducting gap reaches its maximum
 and vanishes respectively~\cite{Gallais_PRB05B}. Here ERS is performed on HgBa$_{2}$CuO$_{4+\delta }$
 (Hg-1201) one of the most simple structure which takes a pure tetragonal symmetry
 with only one copper oxide plane per unit cell. By a judicious choice of the
 laser excitation lines we have enhanced the ERS signal allowing a direct view
 of the electronic properties of this cuprate without invoking \textit{ad hoc}
 phonon subtraction procedures usually used in cuprate systems.

Our results demonstrate that the superconducting state on the underdoped side is also
not normal, and cannot be described within the standard \textit{d}-wave
BCS model. We show that a minimal description of this state requires two energy
scales and two quasiparticle dynamics. One scale is associated with the nodal
region, and decreases with decreasing doping, while the second is
associated with the antinodal region, and increases with decreasing doping.
 Close to the nodes the quasiparticle spectral weight remains substantial
 even at low doping in contrast to the antinodes where it drastically decreases.
 As a consequence i) we show that the superconducting order parameter is
 controlled by two parameters (the slope of the gap function at the nodes and
 the antinodal gap maximum) instead of one as expected from a standard
 \textit{d}-wave gap, ii) we find a strong momentum dependence of
 the quasiparticles spectral weight in the superconducting state of
 the underdoped regime, and iii) we finally establish,
 using a new Raman sum-rule, a
 simple
 relationship between the superfluid
 density and the low-energy Raman scattering associated with nodal
 physics, suggesting that the Fermi liquid renormalization
 of the current (vector quantity) and stress tensor (which
 transforms as a product of currents) have similar doping dependence.
Our new experimental results place strong constraints on current
theories of the high temperature superconductivity phenomenon.

\section*{EXPERIMENTAL RESULTS}

The ERS measurements (see Methods section for details)
have been performed on selected, as-grown, Hg-1201 single crystals from
several batches with different doping levels extending from the slightly over-doped (OD) to the underdoped (UD)
regime: ($T_c$ = 92K(OD), 95K(OP), 89K(UD), 78K(UD) and 63K(UD)).

ERS is an energy probe, but also a momentum probe which
allows us to select different parts of the Fermi surface by choosing combinations of incident
and scattered light polarizations~\cite{Gallais_PRB05B}. The NR and ANR have been explored
using cross polarizations parallel to the Cu-O bond directions (B$_{2g}$) and at 45$^o$ from
them (B$_{1g}$), respectively.
%
All spectra have been corrected for the spectral response of the
spectrometer and for the Bose-Einstein factor. They are thus
proportional to the imaginary part of the Raman response function
$\chi ^{\prime \prime}(\omega)$.

\begin{figure}[tbh]
\begin{center}
\includegraphics{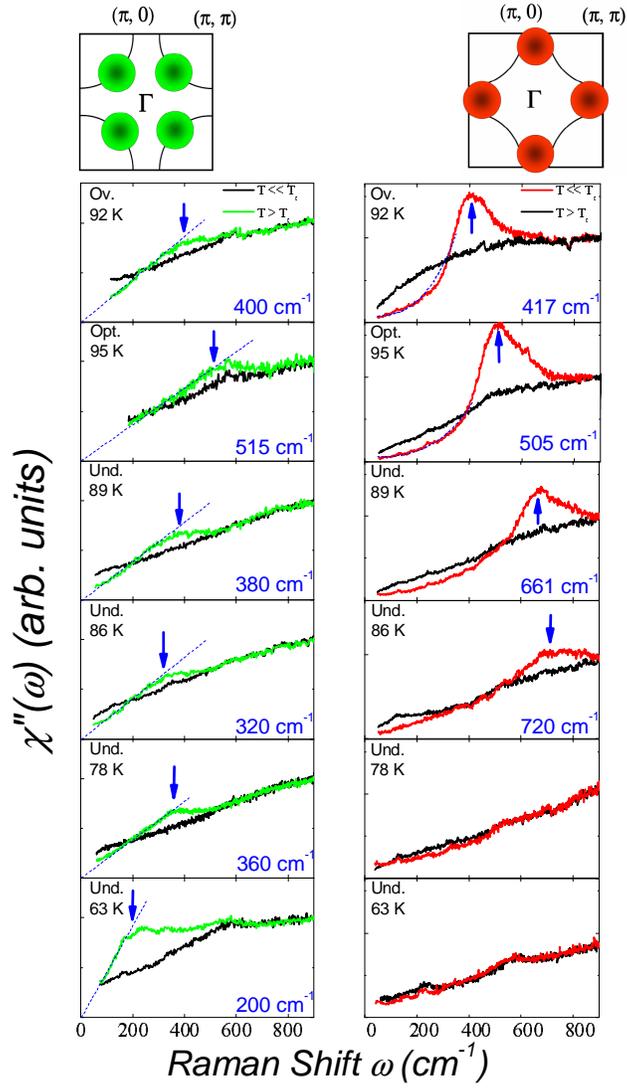} 
\end{center}\vspace{-5mm}
\caption{Raman responses in the NR (B$_{2g}$) and ANR (B$_{1g}$) as function of doping.
The dashed lines on (B$_{2g}$) (respectively (B$_{1g}$)) spectra show the linear (resp. cubic) frequency dependences of the nodal
(resp. antinodal) Raman responses. The arrows indicate the position of the superconducting peak maxima.}
\label{fig:B1gB2gResponses}
\end{figure}

\vskip .5cm
%
%
Figure~\ref{fig:B1gB2gResponses} displays the NR (B$_{2g}$) and ANR (B$_{1g}$) Raman response
functions in both the normal and superconducting states of the Hg-1201 single crystals, at
various doping levels.
At optimal doping ($T_c$=95 K), the electronic Raman continua for both the ANR and NR exhibit
a redistribution of spectral weight from  energies
lower than 400 cm$^{-1}$ to higher energy, when going from the normal state to the
superconducting state. This redistribution is more pronounced for the ANR than for the NR.
 At low energy (below 400 cm$^{-1}$), the ANR superconducting continuum exhibits a cubic
 frequency dependence with a well-marked superconducting pair breaking peak, indicated by
 an arrow on Fig.\ref{fig:B1gB2gResponses}, at a frequency $\hbar\omega_{AN} \simeq$ 505 cm$^{-1}$
 ($\simeq$ 8$k_BT_c$). In contrast, the superconducting spectrum in the NR displays a
 linear frequency dependence up to 400 cm$^{-1}$, as well as a weaker signature of the
 pair-breaking peak (arrow) close to the same frequency
 $\hbar\omega_{N} \simeq \hbar\omega_{AN}$ than for the ANR spectrum (but with a half width
 at half maximum reduced approximately by a factor of two). The Ramanresponse at optimal
 doping is thus characterized by a single energy scale $\hbar\omega_{AN} \simeq\hbar\omega_{N}$
 associated with the pair-breaking peak, and all the features described above are consistent with those
expected for a \textit{d}-wave superconductor~\cite{Devereaux_PRB95} with a maximum value $\Delta_m$ of the
superconducting gap given by $2\Delta_{m} = \hbar\omega_{AN}$ (see below).
Our results for one overdoped sample (spectra at the top of Fig.~\ref{fig:B1gB2gResponses}) can also be
interpreted in terms of a single energy scale.

In contrast, as doping is decreased below the optimal one, the evolution of the Raman spectra
in the superconducting phase becomes strikingly different in the ANR and in the NR.
 As the doping level (and $T_c$) is reduced, the characteristic energy of the antinodal
 peak (indicated by an arrow on the right pannel of Fig.~\ref{fig:B1gB2gResponses})
 increases.
Simultaneously, the intensity of this peak rapidly decreases as $T_c$ decreases, and
finally disappears in the vicinity of $T_c$=78K. In contrast, the characteristic energy
scale of the spectrum in the NR (which we take to be the frequency $\omega_{N}$ of the maximum
observed in $\chi^{\prime \prime}(\omega)$) follows $T_c$. Furthermore, contrarily to the antinodal peak,
 the nodal peak persists down to the lowest doping that we have studied ($T_c$=63K). We note that similar
 observations have been reported previously on other cuprate materials,
 such as Bi$_2$Sr$_2$CaCu$_2$O$_{8+\delta}$ (Bi-2212),
 Bi$_2$Sr$_{2-x}$La$_x$CuO$_{6+\delta}$ (Bi-2201), YBa$_2$Cu$_3$O$_{7-\delta}$
 (Y-123) and La$_{2-x}$Sr$_x$CuO$_4$ (LSCO)~\cite{Venturini_JPCS01, Sugai_PRB03}.
This demonstrates that the electronic Raman response in the underdoped regime involves
two distinct energy scales, with opposite doping dependence. As discussed below,
this is inconsistent with a simple BCS \textit{d}-wave description~\cite{Devereaux_PRB95}.

\begin{figure}[tbh]
\begin{center}
\includegraphics{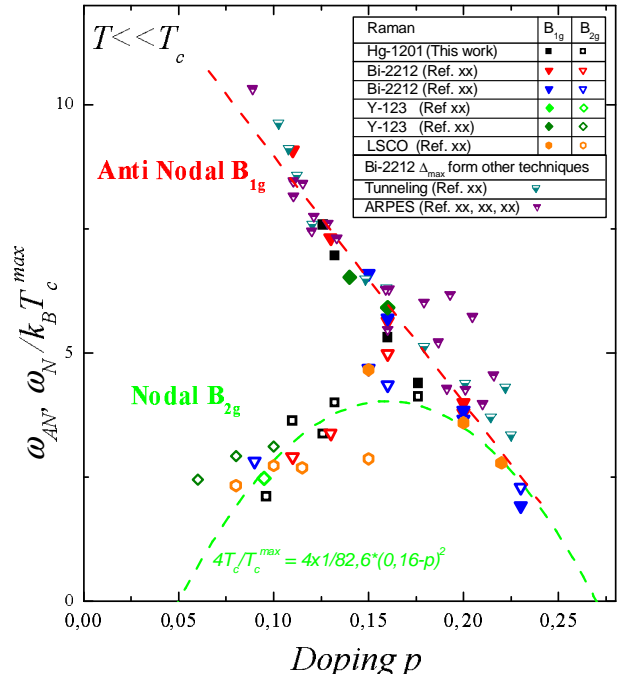} 
\end{center}\vspace{-5mm}
\caption{Universal doping dependence of the ratios $\hbar\omega_{AN}/k_BT_c^{max}$ and $\hbar\omega_{N}/k_BT_c^{max}$ obtained from the B$_{1g}$ and B$_{2g}$ superconducting Raman peaks. The ratios $2\Delta/k_BT_c^{max}$ deduced from ARPES coherent peak in the ANR ~\cite{Norman_Nature98, Shen_Science98, Campuzano_PRL99} and from tunneling spectroscopies ~\cite{Miyakawa_PRL98, Miyakawa_PRL99, deWilde_PRL98} have also been reported.}
\label{fig:B1gB2gscales}
\end{figure}

In order to substantiate further this point, we have plotted in Fig.~\ref{fig:B1gB2gscales}
 the characteristic ratios $\hbar\omega_{AN}/k_BT_c^{max}$ and
$\hbar\omega_{N}/k_BT_c^{max}$ obtained for several different families
 of cuprates by different groups~\cite{Sugai_PRB03, Chen_PRB97, Venturini_JPCS01},
 as a function of doping at a fixed temperature well below $T_c$.
 The doping value p is inferred from $T_c$ using Tallon's
equation~\cite{Presland_PhysicaC91}: $1-T_c/T_c^{max} = 82.6\,(p-0.16)^2$,
 with $T_c^{max}$ the critical temperature at optimal doping.
 Fig.~\ref{fig:B1gB2gscales} reveals that these ratios have a universal
 dependence on doping. For underdoped compounds, two distinct scales are
 present, with the two ratios behaving in opposite manners as a function
 of doping, while a unique energy scale and doping dependence is
recovered at optimal doping and in the overdoped regime.

\section*{INCONSISTENCY WITH A SIMPLE BCS MODEL}

Let us now analyze these results using the simplest possible framework,
 that of a BCS superconductor with a
\textit{d}-wave gap function of the form $\Delta_k =
 \Delta_m\cos(2\phi)$ (where $\phi$ is
the angle associated with momentum ${\bf k}$ on the Fermi surface).
 The Raman response would then read~\cite{Klein_PRB84,Devereaux_PRB95}:
\begin{equation}
\chi^{\prime \prime}_{AN, N}(\omega)
= \frac{2\pi N_F}{\omega}\left\langle \frac{(\gamma^{AN,
N}(\phi))^2\Delta_m^2\cos^2(2\phi)}{\sqrt{\omega^2 - 4\Delta_m^2\cos^2(2\phi)}}\right\rangle _{FS}
\end{equation}
$\gamma^{AN, N}(\phi)$ is the Raman vertex associated with each polarization:
$\gamma^{AN}(\phi) = \gamma_{B_{1g}}\cos(2\phi)$
while $\gamma^{N}(\phi)= \gamma_{B_{2g}} \sin(2\phi )$ , and $\left\langle ... \right\rangle_{FS}$
denotes a Fermi-surface average. This predicts a sharp pair-breaking peak (corresponding to a divergence of this
expression) in the B$_{1g}$ (ANR) geometry at $\omega=2\Delta_m$, and a weaker singularity in the B$_{2g}$
(NR) geometry at the same frequency scale
(furthermore, within the simplest BCS formula above, the $B_{2g}$ (NR) response displays a
maximum at a somewhat lower energy than the peak in the $B_{1g}$ (ANR)
channel, but both are governed by one energy scale, that of the maximal gap $\Delta_m$).
This is roughly consistent with the experimental observations at optimal
doping, but completely fails to account for the underdoped spectra (for which the observed peaks become clearly
distinct in each polarization and have opposite doping dependence).

This clearly demonstrates that one or both of the following
assumptions become invalid in the underdoped regime: i)
non-interacting BCS quasiparticles ii) a gap function with the
simple form $\Delta_k = \Delta_m\cos(2\phi)$ characterized by a
single energy scale. Moving away from assumption i) requires
taking into account, in the framework  of the  Landau theory of
interacting quasiparticles, the spectral weight $Z_k$ of these
quasiparticles, in general smaller than one and k-dependent, as
well as the  Fermi-liquid vertex  $\Lambda_k$ describing the
 interaction of the quasiparticles with external perturbations.
 This leads to:
\begin{equation} \chi^{\prime \prime}_{AN, N}(\omega) = \frac{2\pi N_F}{\omega}\left\langle
\frac{(Z\Lambda)_k^2(\gamma^{AN, N}_k)^2\Delta_k^2}{\sqrt{\omega^2 - 4\Delta_k^2}}\right\rangle _{FS}
\label{RamanFL}
\end{equation}
in which a general gap-function has also been taken into account. This expression contains two
unknown functions of momentum on the Fermi-surface however:
$(Z\Lambda)_k$ and $\Delta_k$, and constructing on this basis a phenomenological
analysis (to which we shall come back later in this paper) requires further insight.

\section*{LOW-ENERGY NODAL EXCITATIONS, SUM-RULE AND RELATIONSHIP TO THE SUPERFLUID DENSITY}

\begin{figure}[tbh]
\begin{center}
\includegraphics{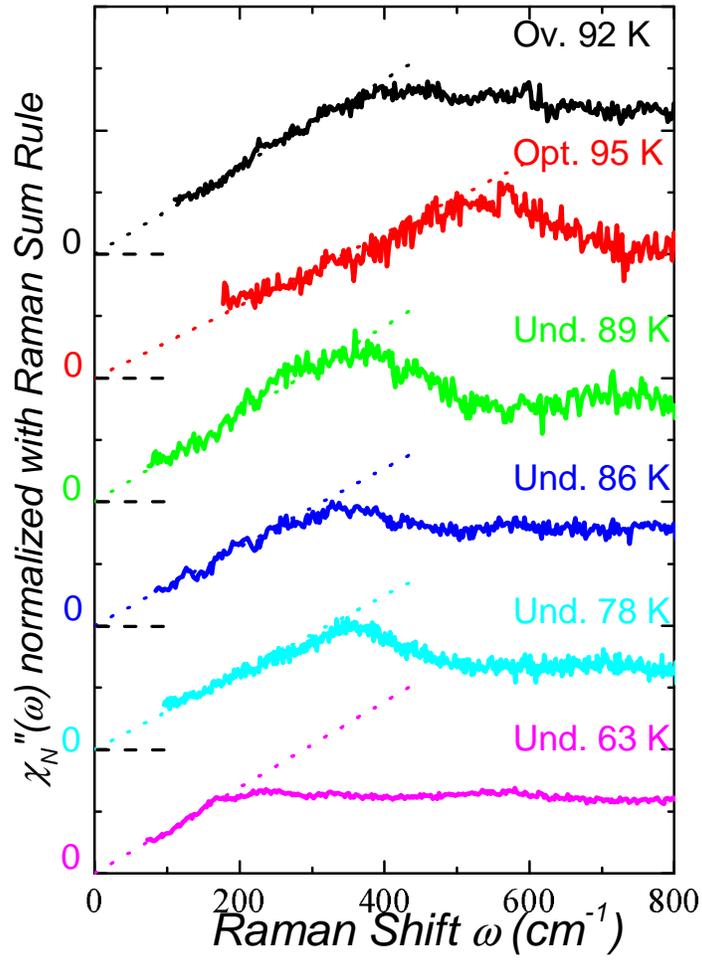} 
\end{center}\vspace{-5mm}
\caption{Normalized Raman response functions with respect to the sum rule.
A weak linear background coming from spurious luminescence for intermediate doping,
independent of the scattering geometry and excitation lines has been subtracted
from raw data before performing the normalization
(note that without this subtraction the final result is qualitatively similar,
i.e the low energy slopes of the normalized nodal Raman response are found to be
doping independent).}
\label{fig:B2gRenorm}
\end{figure}

In order to gain such insight, we have focused on the low-energy part of the Raman spectra,
which is controlled by the properties of the nodal quasiparticles. The B$_{2g}$ geometry
is particularly significant in this respect, since the largest low-energy response is
 obtained in this geometry and the NR is directly sampled. Figures~\ref{fig:B1gB2gResponses}
 and~\ref{fig:B2gRenorm} demonstrate that a linear dependence on frequency is found
in this geometry, for several dependent doping levels. This is expected from the above formula,
 which yields:
%
$\chi^{\prime \prime}_{N}(\omega\rightarrow 0) = \gamma_{B_{2g}}^2
\frac{\pi^2 N_F}{2v_{\Delta}}(Z\Lambda)^2_N\omega + ...$.
In this expression, $v_{\Delta} = \frac{d\Delta}{d\phi}|_N$ is the slope of the gap
 function at the nodes, and $(Z\Lambda)_N$ is the value of $(Z\Lambda)_k$ at the node.
Hence, a study of the doping-dependence of the slope of the nodal (B$_{2g}$) response,
 allows determining, in principle, the important parameter
 $\alpha = \frac{N_F}{v_{\Delta}}(Z\Lambda)_N^2$ associated with nodal physics.

\begin{figure}[tbh]
\begin{center}
\includegraphics{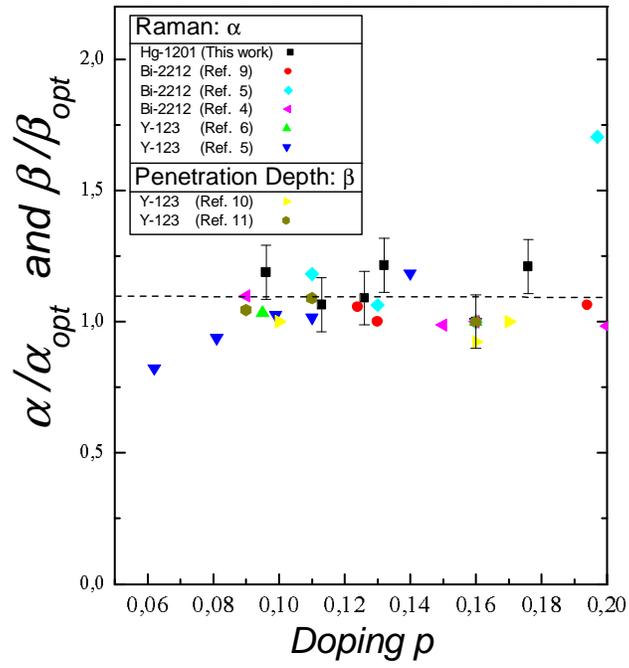} 
\end{center}\vspace{-5mm}
\caption{Doping dependence of the slope of the nodal Raman response
$\alpha = \frac{N_F}{v_{\Delta}}(Z\Lambda)_N^2$, normalized to the optimal doping one ($p=0.16$).
The Fermi liquid parameter $\beta = \frac{N_F}{v_{\Delta}}(Z\Lambda_{\rho})_N^2$
extracted form the temperature dependence of the penetration depth ~\cite{Panagopoulos_PRL98, Bonn_CJP96}, is also shown.
$\alpha$ and $\beta$ are both found to be doping independent in the range $(p=0.09-0.020)$.}
\label{fig:B1gB2gPDbilan}
\end{figure}

In order to compare the slopes of samples with different doping levels however,
 the Raman spectra must be properly normalized. To achieve this goal, we have
established a theoretical sum-rule for the Raman intensity of a weakly doped Mott
 insulator, which reads, at low doping levels $p$: $\int_0^\Omega \chi^{\prime \prime}
 (\omega)\omega d\omega = Cp$. We have derived this expression (see Methods section below)
 by starting from a one-band t-J model, in which case the upper cutoff $\Omega$ can be taken
 to infinity, and the constant C depends on the hopping amplitudes and exchange constant:
 $C(t,t^\prime,J)$. We expect this expression to be general for a weakly doped Mott insulator,
 provided the upper cutoff is taken to be of the order of the bare bandwidth and significantly
lower than the scale U associated with the upper band. We have normalized the Raman B$_{2g}$
 data using this sum-rule for our own data on Hg-1201. This is shown in Fig.~\ref{fig:B2gRenorm},
 and reveals that in the doping range of interest the slope is doping- independent.
In order to point out the universality of this phenomena, we have applied the same
Raman sum-rule to a number of spectra previously published in the literature for
Y-123 and Bi-2212~\cite{Venturini_JPCS01, Chen_PRB97, Sugai_PRB03, Hewitt_PRB02}.
%
The results for the slope $\alpha$ as a function of doping are displayed on
Fig.~\ref{fig:B1gB2gPDbilan}. It is seen that the slope is essentially doping-independent
 for all doping levels between optimal doping (p=0.16) down to p=0.09.
 For smaller doping levels, the Y-123 data (two data points available)
 suggest that $\alpha$ may end up decreasing at very low doping.

This doping-independence of the Raman B$_{2g}$ (NR)
slope over an extended range of doping levels is very reminiscent of the
behavior of the slope of the linear term in the temperature-dependence of the penetration
 depth (or superfluid density)~\cite{Bonn_CJP96, Panagopoulos_PRL98}:
$\rho_s(T)=\rho_s(0)-\beta T+\cdots$. This quantity is also associated with the physics of
 nodal quasiparticles, and given by almost the same expression~\cite{Lee_PRL97, Durst_PRB00}
%
%
$\beta \propto \frac{N_F}{v_{\Delta}}(Z\Lambda_{\rho})_N^2$,
except for the fact that the Fermi liquid parameter $\Lambda_{\rho}$ corresponds
to a different angular-momentum
channel than in the case of Raman. The comparison made in Fig.~\ref{fig:B1gB2gPDbilan} between these
two quantities shows excellent agreement, hence establishing a previously unforeseen relation
between Raman and penetration depth measurements.
We also note that very recent experiments~\cite{Broun_Condmat0509223} do suggest deviations from
$\beta = \rm{const}.$ at very low doping levels, in agreement with the two highly underdoped Y-123 data
 points in Fig.~\ref{fig:B1gB2gPDbilan}.
 As pointed out in Ref.~\cite{Lee_PRL97}, the independence of the slope $\beta$ of the superfluid
 density upon doping, in an extended
 range, is a serious difficulty for the simplest (``vanilla'') version of RVB
theory~\cite{Anderson_Science87}, in either its auxiliary
boson~\cite{Kotliar_PRL88} or wave function
formulations~\cite{gros} \cite{Anderson_JPCM04}. Our observations
reveal a similar problem in connection with Raman scattering.
%

Having established the doping-(in)dependence of the $\alpha = \frac{N_F}{v_{\Delta}}(Z\Lambda)_N^2$
 parameter associated with nodal physics, we can make further progress in disentangling the relative
 effects of i) quasiparticle dynamics and ii) the form of the gap function on the Raman spectra.
 Let us first assume that the gap function has the simple form:
$\Delta_k = \Delta_m \cos(2\phi)$ characterized by a single energy scale (the maximum gap), and such that
$v_{\Delta}=2\Delta_m$. Since the behavior of the ANR (B$_{1g}$) peak implies that $\Delta_m$
increases as doping is reduced (Fig.~\ref{fig:B1gB2gscales}), the doping-independence of
$\alpha$ would then imply that $(Z\Lambda)_N$
must correspondingly increase as $p$ is reduced.
This behavior
is highly unlikely to apply for a doped Mott insulator,
since it would correspond to
a reinforcement of the
quasiparticle spectral weight as doping is reduced.
 Furthermore, ARPES data
and theory
do suggest that the nodal quasiparticle spectral weight decreases
as p is reduced~\cite{Yoshida_PRL03, Shen_Science05}. Hence, we
are led to suspect that the slope of the gap at the nodal points,
$v_{\Delta}$ , must decrease as the doping level is reduced (in
order to keep $\alpha$ constant) and hence does not track the
maximum gap $\Delta_m$. This, of course, has far reaching
consequences, namely: that a pure $\cos(2\phi)$ form does not
hold in the underdoped regime, and that two energy scales
characterize this regime (as already foreseen from the above
analysis in Fig.~\ref{fig:B1gB2gscales}). Indeed, this has been
previously suggested from ARPES experiments~\cite{Mesot_PRL99,
Borisenko_PRB02R}, indicating that the superconducting gap
function may change from a ``V-shape'' to a ``U-shape'' as the
doping level is reduced.

\section*{PHENOMENOLOGICAL DESCRIPTION}

\begin{figure}[tbh]
\begin{center}
\includegraphics{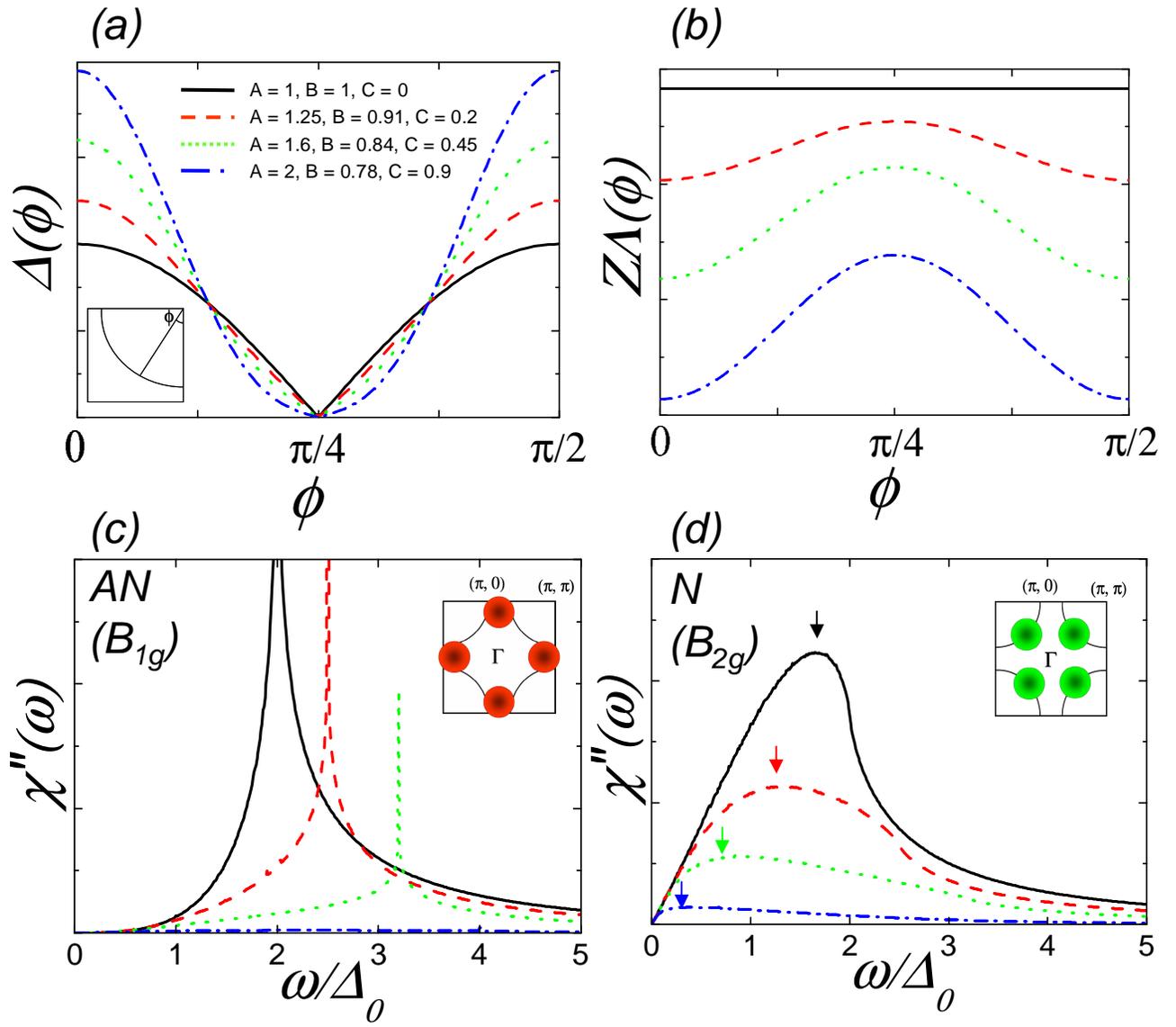} 
\end{center}\vspace{-5mm}
\caption{
(a) Momentum dependence of the phenomenological superconducting gap function: $\Delta(\phi) = A\times \Delta_{0}(B\cos 2\phi+(1-B)\cos 6\phi)$. A is related to the maximum amplitude of the superconducting gap, and B to its deviation to the pure \textit{d}-wave.
(b) Momentum dependence of the phenomenological $Z\Lambda$ function: $Z\Lambda(\phi) = \sqrt{v_\Delta}\times (1-Ccos^2(2\phi))$, where $v_\Delta = \frac{d\Delta}{d\phi}|_{\phi=\frac{\pi}{4}} =\Delta_m(8B-6)$.
(c) Calculated ANR (B$_{1g}$) response.
(d) Calculated NR (B$_{2g}$) response function. Arrows indicate the position of the maximum.
}
\label{fig:fit_pheno}
\end{figure}

We now return to the general expression for
the Raman response based on Fermi liquid considerations, and explore whether reasonable momentum- and
doping-dependence of the two functions $Z\Lambda(\phi)$ and $\Delta(\phi)$ (associated respectively with
quasiparticle physics and with the superconducting gap) can describe our data.
We use the shape of the gap function (consistent with \textit{d}-wave symmetry) proposed from ARPES data in
Ref.~\cite{Mesot_PRL99}: $\Delta(\phi) = \Delta_{m}(B\cos 2\phi+(1-B)\cos 6\phi)$, where
$0 \leq B \leq 1$.
This form is characterized by two scales: the (antinodal) maximum gap
$\Delta_{m}$ and the nodal slope $v_\Delta = \Delta_m(8B-6)$.
In a first attempt, we have taken $Z\Lambda$ to be
$\phi$-independent (uniform along the Fermi surface). The
parameters $\Delta_m$ and $B$ were varied, as in
Ref.~\cite{Mesot_PRL99}, and the values of $Z\Lambda$ chosen such
as to keep the ratio $(Z\Lambda)_N^2/v_\Delta$ constant, for
consistency with our experimental observation on the doping-
independence of the slope $\alpha$. We found that the resulting
spectra (not shown) could capture only part of our experimental
observations: the B$_{1g}$ peak does move to higher energy but
does not loose intensity quickly enough, and the B$_{2g}$
response becomes very flat as $v_\Delta/\Delta_m$ is reduced, but
its maximum does not clearly recede towards lower energy. Hence,
a variation of $Z\Lambda(\phi)$ along the Fermi surface is
clearly needed, which we have chosen as displayed in
Fig.~\ref{fig:fit_pheno}b (chosen again in such a way that
$(Z\Lambda)^2_N/v_\Delta$ is kept constant). The four curves
colors in Fig.~\ref{fig:fit_pheno} are meant to correspond,
qualitatively, to the overdoped (BCS), optimally doped, slightly
and strongly underdoped regimes. The resulting spectra are
displayed in Figs.~\ref{fig:fit_pheno}c-d for both B$_{1g}$ and
B$_{2g}$ geometries. All the  main
qualitative  aspects of our experimental observations are
captured by this simple phenomenological model,
namely: the opposite variation of the nodal and antinodal peaks
with energy, the constant nodal slope, and the drastic
suppression of the antinodal peak due to the incoherence of the
antinodal quasiparticles (consistent with Fermi surface
``arcs''~\cite{Norman_Nature98}).

\section*{COMPARISON TO OTHER EXPERIMENTS}

Finally, we discuss whether other experimental probes support the
existence of two energy scales in the superconducting state of
the underdoped cuprates.

Let us first note that the two basic building blocks of the phenomenological analysis
developed above are consistent with available information from ARPES experiments.
The leading edge in the ANR (corresponding to the higher energy scale) was shown to
increase as the doping level is reduced~\cite{Feng_Science00}. Evidence for a second energy scale
associated with the nodal region, and varying in an opposite manner with doping,
has been suggested from ARPES in Refs.~\cite{Mesot_PRL99, Borisenko_PRB02R}, with a corresponding
change of the gap function as discussed above. More extensive ARPES
studies of the leading edge in the NR are clearly needed.
ARPES also reveals that the (gapped) antinodal quasiparticles in the superconducting state
quickly loose spectral weight and eventually become incoherent as doping is
reduced~\cite{Shen_Science05, Ding_PRL01}, while nodal quasiparticles loose spectral weight as
doping is reduced but maintain coherence down to low doping levels~\cite{Yoshida_PRL03, Shen_Science05}.

It was suggested from thermal conductivity measurements on Y-123~\cite{Sutherland_PRB03} that the
nodal slope of the gap, $v_\Delta$ increases as the doping level is reduced, which would
seem in contradiction with our observations. However,
closer examination~\cite{Taillefer_private}
reveals that only one data point is available in the doping range of interest
here, and more experiments are clearly needed.

Tunneling, while not being a momentum- selective probe like Raman or ARPES,
exhibits a set of distinct superconducting line shapes as a function of doping.
The main peak in these spectra moves to higher energies as doping is reduced from
optimal doping~\cite{McElroy_Nature03,McElroy_PRL05}
(as also seen from break-junction measurements~\cite{Miyakawa_PRL98,Miyakawa_PRL99}),
and at the same time broadens considerably. This is consistent with the behavior of the
larger energy scale, and with the rapid loss of coherence of antinodal quasiparticles. Fourier transform tunneling has shown that antinodal decoherence is closely related to a local charge order, at least in surface~\cite{McElroy_PRL05}. This is consistent with the loss of the electronic Raman response which is sensitive to the charge dynamics.
In contrast, the observation of quasiparticle interferences from tunneling~\cite{Hoffman_Science02_2}
provides strong evidence for the coherence of nodal quasiparticles.
There are also some recent indications that the lower energy scale might show
up as a low-voltage shoulder in the tunneling spectra of underdoped samples~\cite{McElroy_PRL05}.
There are also some previous indications in the literature that the
lower energy scale shows up as the
gap edge in the Andreev reflection spectra of underdoped samples~\cite{Deutscher_Nature99}.

\section*{CONCLUSION}

In summary, electronic Raman scattering clearly demonstrates the existence
of two distinct energy scales in the superconducting state of underdoped cuprates,
depending on doping in opposite ways.
Here, we have suggested that these two scales are associated with the nodal slope
of the gap function, and its maximum, respectively.
%
%
Correspondingly, our experiments reveal two different dynamical
properties of the quasiparticles in these two regions. Hence, the
dichotomy between the coherence of nodal quasiparticles and the
incoherence of antinodal ones, which is well established in the
normal state, persists in the superconducting state in the
underdoped regime. A theoretical analysis using a new sum-rule
reveals that the superconducting state cannot be described within
the standard BCS-Migdal Eliashberg theory applied with a pure
$(\cos k_x-\cos k_y)$ gap. We have also established a new
relation between the behavior of the low-energy Raman response
and that of the temperature-dependence of the superfluid
stiffness, both associated with nodal physics. All these
observations emphasize the key importance of the momentum space
differentiation between nodal and antinodal regions in both the
normal and superconducting state
in the proximity to the Mott transition. The physics of momentum
space differentiation can have multiple origins and has been
stressed  early on by many authors\cite{pines,millis}.  Modern
theoretical approaches, such as cluster perturbation theory,
functional renormalization groups, and cluster extensions of
Dynamical Mean Field Theory have shown to be able to  capture this
physics in both the weak coupling and strong coupling regimes of
the Hubbard model~\cite{katanin04, Senechal_PRL04,Civelli_PRL05}.
Our results  provide further tests of these methods and should
stimulate further development of these approaches,
aiming in particular at a quantitative evaluation of the ERS response.

\section*{METHODS}
\subsection*{Details of the experimental procedure}

The Hg-1201 single crystals have been grown by the flux method. The detailed procedure of the
crystal growth is described elsewhere~\cite{Bertinotti_PhysicaC96}.
The magnetically measured transition widths are less than 5K for all samples.
The ERS experiments have been carried out using a triple grating spectrometer
(JY-T64000) equipped with a nitrogen cooled CCD detector.
The crystals were mounted in near-vacuum (10$^{-6}$mbar) on the cold finger
of a liquid helium cryostat.
Rotation of the crystal was achieved in situ. Red (1.9 eV) and Green (2.4 eV) excitation lines
have been used for probing the ANR and NR Raman response functions, respectively,
in order to magnify the ERS signal. It has been experimentally
shown that resonance effects occur for these lines and are probably due to the matching of the excitation
energies to the inter-band transitions of Hg-1201~\cite{LeTacon_PRB05}.
The spectra presented here have been taken below (T = 10 K, corrected from laser heating), and above $T_c$ (T = 100 K for 92K(OD), 95K(OP) and 89K(UD) crystals, and T = 89 K for 78K(UD) and 63K(UD) crystals).

\subsection*{Theoretical sum-rule}

The absolute value of the Raman intensity in
samples with different doping levels or composition is difficult to
ascertain experimentally, because it depends on geometric factors
which are hard to measure,
such as the  penetration volume of the electromagnetic field into
the sample.
To circumvent this problem, and to
make contact with the physics of the Mott insulator, we use a
sum rule to normalize the Raman intensity.
In Ref.~\cite{Freeriks_PRL05}, such a sum rule was derived for the
Hubbard model. This cannot be used to interpret our experiment
however, because it involves the contribution
of the Raman intensity from the upper Hubbard band, which
cannot be determined experimentally since it occurs at energies
where other interband transitions take place.
For this reason, we have extended the method of Ref.~\cite{Freeriks_PRL05} and
derived a new low-energy sum rule for the B$_{2g}$
Raman intensity, directly for the $t-J$ model (for which the upper
Hubbard band has been projected out).
This sum-rule relates the low energy Raman intensity to
the expectation value of a composite operator:
\begin{equation}\nonumber
\int d\omega\, \omega \chi''(\omega) = \frac{\pi}{2}< [M,[H,M^\dagger]]> =
\sum_{ijkl}\sum_{\alpha\cdots\delta'}
T_{ijlm}^{\alpha\alpha'\cdots\delta\delta'}
<X_{\alpha\alpha'}(i)X_{\beta\beta'}(j)X_{\gamma\gamma'}(l)
X_{\delta\delta'}(m)>
\end{equation}
In this expression, $X_{\alpha\alpha'}(i)$ is the Hubbard
operator at site $i$, and the Greek indices $\alpha\cdots\delta'$
run over the three possible atomic states $0,\uparrow,\downarrow$.
$M$ is the B$_{2g}$ Raman operator (stress tensor) $M=\sum_{{\bf
k}\sigma} \sin {k_x } \sin{ k_y}  X_{\sigma 0 }({\bf k}) X_{0
\sigma}({\bf k})$ and $H$ is the $t-J$ Hamiltonain (containing a
hopping term and a superexchange interaction). We have calculated
the tensor $T_{ijlm}^{\alpha\alpha'\cdots\delta\delta'}$
appearing on the r.h.s of this equation. For a hopping term
involving nearest ($t$) and next-nearest neighbor ($t'$),
assuming unbroken spatial and time reversal symmetry,
we have checked that all non-zero contributions involve at least
two fermionic operators (i.e. $X_{\sigma 0}$ or $X_{0\sigma}$)
living on different sites.
In the absence of translational
symmetry breaking
contributions involving exactly two such operators are
proportional to doping. We have checked this from an explicit
evaluation using e.g. slave-boson or dynamical mean-field
theories. Details will be published elsewhere. Hence, the
B$_{2g}$ Raman intensity for the $t-t'-J$ model reads:
$\int_0^\infty d\omega\, \omega \chi''(\omega) =
C(t,t',J)\,p+\cdots$ at low doping levels, justifying the
normalisation used above. A similar observation follows from a different
sum rule \cite{Shastry_PRL90}.



\section*{ACKNOWLEDGEMENTS}

We are grateful to: S. Biermann, N. Bontemps, S.V. Borisenko, P. Bourges, M. Cazayous, R. Combescot, G. Deutscher, T.P. Devereaux, K. McElroy, P. Monod, M. Norman, C. Panagopoulos, Z.-X. Shen, and L. Taillefer for useful discussions.
 This research was supported by CNRS, by Ecole Polytechnique and by the "Chaire Blaise Pascal de la
Fondation de l'Ecole Normale Sup\'erieure et de la r\'egion Ile de France".

\newpage

\newpage

\newpage

\newpage

\newpage

\end{document}